# The impact of classical bulges on stellar bars and boxy–peanut–X features in disc galaxies

Rachel Lee McClure ,[1]★ Angus Beane,[2] Elena D'Onghia,[1] Carrie Filion [3] and Kathryne J. Daniel [4]

[1]*Department of Astronomy, University of Wisconsin–Madison, Madison, WI 53706, USA*
[2]*Center for Astrophysics | Harvard & Smithsonian, Cambridge, MA 02138, USA*
[3]*Center for Computational Astrophysics, Flatiron Institute, New York, NY 10010, USA*
[4]*Department of Astronomy & Steward Observatory, University of Arizona, Tucson, AZ 85721, USA*

## ABSTRACT

Galactic bars and their associated resonances play a significant role in shaping galaxy evolution. Resulting resonance-driven structures, like the vertically extended boxy–peanut–X (BPX) feature, then serve as a useful probe of the host galaxy's history. In this study, we quantify the impact of a classical bulge on the evolution of the bar and the growth of bar-resonance structures. This is accomplished with a suite of isolated *N*-body disc galaxy simulations with bulge mass fractions ranging from 0 per cent to 16 per cent of the disc mass. We apply frequency analysis to the stellar orbits to analyse the variations in resonance structure evolution. Our findings indicate that a more massive initial bulge leads to the formation of a stronger and more extended bar and that each bar drives the formation of a prominent associated BPX through resonance passage. In this work, we present evidence that the formation of a BPX is driven by planar, bar-supporting orbits evolving through interaction with horizontal and vertical bar resonances. More orbits become vertically extended when these resonances intersect, and the rate of the orbits passing through resonance is moderated by the overall fraction of vertically extended orbits. A significant bulge stabilizes the fraction of vertically extended orbits, preventing sudden resonance-induced changes. Crucially, neither sudden resonance intersection nor prolonged resonance trapping is required for BPX formation.

## 1 INTRODUCTION

Central bars are common features with observations revealing them in more than 60 per cent of nearby disc galaxies (Eskridge et al. 2000; Elmegreen, Elmegreen & Hirst 2004; Sheth et al. 2008; Aguerri, Méndez-Abreu & Corsini 2009; Masters et al. 2010) and the presence of stellar bars as early as *z* ∼ 3 (Bland-Hawthorn et al. 2023; Guo et al. 2023). Disc galaxies vary widely in observed morphology, with many stellar bars accompanied by a central, bulkier mass component broadly categorized as a bulge. These bulges are split into classical bulges dominated by random motions that form throughout the hierarchical assembly process and pseudobulges that include rotationally supported bulge-like structures thought to form through secular resonant processes (Kormendy & Illingworth 1982; Kormendy & Kennicutt 2004; Kautsch 2009; Kormendy & Barentine 2010; Ness et al. 2014).

Observational results reveal that bulge fraction varies across disc galaxies and that not all barred galaxies have a central bulge with ∼20–70 per cent of local massive disc galaxies observed to be pure discs (Kormendy & Kennicutt 2004; Kautsch 2009; Kormendy et al. 2010). Early-type discs are likely to host a significant classical bulge and bulge-to-total luminosity ratios range from ∼2 per cent to 40 per cent in later-type disc galaxies (Aguerri, Balcells & Peletier 2001; Kormendy & Kennicutt 2004; Eliche-Moral et al. 2006). In general, bulges are expected to be long-lived and common as cosmological simulations reveal that central stellar bars are generally accompanied by a central bulge contributing 40–50 per cent of the stellar mass (Du et al. 2020).

Many stellar bars are observed to have a vertically thick stellar region emerging from the centre of the bar that takes the shape of a boxy–peanut bulge or even a sharp X-shape, collectively referred to as a boxy–peanut–X (BPX) feature (e.g. Whitmore & Bell 1988; Combes et al. 1990; Raha et al. 1991; Athanassoula 2005; Bureau et al. 2006; Laurikainen & Salo 2016). Over 40 per cent of local massive disc galaxies are observed with BPX bulges (Burbidge & Burbidge 1959; Shaw 1987; Lütticke, Dettmar & Pohlen 2000a, b; Laurikainen et al. 2011, 2014) with a significant increase in the number of galaxies with BPX when the stellar mass exceeds $\log(M_\star) \gtrsim 10.4$ (Marchuk et al. 2022).

These structures are typically classified as pseudobulges due to their dominant rotational kinematics and their formation via internal secular processes rather than hierarchical mergers (Kormendy & Illingworth 1982; Kormendy & Kennicutt 2004; Ness et al. 2014). Observations reveal a wide variety in observed BPX characteristics and bulge fraction across the diverse spread of observed bar parameters (Ciambur & Graham 2016; Erwin & Debattista 2017; Laurikainen & Salo 2017; Savchenko et al. 2017; Tahmasebzadeh et al. 2024). In our own Milky Way, it is found that a classical bulge contributes ≲10 per cent of the total stellar mass, with the kinematics of stars and gas revealing a rotating pseudobulge or BPX feature with distinct metallicity within the central regions of our Galaxy (Dwek et al. 1995; Shen et al. 2010; Li & Shen 2012; Ness et al. 2012, 2013; Bland-Hawthorn & Gerhard 2016; Ciambur, Graham & Bland-Hawthorn 2017; Lucey et al. 2021).

★ E-mail: rlmcclure@wisc.edu

The evolution of a BPX bulge is intrinsically tied to the development of its stellar bar. In galaxy simulations, the BPX structure emerges from the bar either through short-lived asymmetries (Martinez-Valpuesta & Shlosman 2004; Martinez-Valpuesta, Shlosman & Heller 2006; Łokas 2019b) that were initially proposed to be triggered by an underlying firehose gravitational instability (Raha et al. 1991; Merritt & Sellwood 1994; O'Neill & Dubinski 2003; Martinez-Valpuesta & Shlosman 2004; Bournaud, Combes & Semelin 2005; Debattista et al. 2006; Athanassoula 2008a) or through vertical bar-resonance transitions that modify the morphology of stellar orbits, increasing their vertical height from the disc at apoapsis (Combes et al. 1990; Pfenniger & Friedli 1991; Quillen 2002; Quillen et al. 2014; Sellwood & Gerhard 2020).

Bar resonances drive the secular evolution of the disc through the exchange of angular momentum and orbital energy at equilibrium points in the galactic potential (Toomre 1964; Kato 1971; Weinberg 1985; Kormendy 2013). As stars progress through their orbit, bar-resonance interactions occur where stars cross these points in the bar potential within the disc at regular intervals of their orbital period (Combes et al. 1990; Pfenniger & Friedli 1991; Patsis, Skokos & Athanassoula 2002). Resonant interactions alter stellar orbits, shifting their orbital frequencies as they become non-circular and exhibit (at least temporarily) directional frequency ratios inherited from the responsible resonance.

Resonances of the bar are distributed throughout the disc at radial intervals where the azimuthal frequency of the orbits are integer intervals of the difference between the orbital azimuthal frequency and the frequency of the bar orbit ($\Omega_\phi - \Omega_b$). Orbits that span the region out to the last trapped orbit of the bar (sometimes noted as *x1*) can be extremely elongated within the radius of the corotation resonance (Lin & Shu 1964; Patsis et al. 2002) where azimuthal frequency of a circular orbit at that point in the potential is $\Omega_\phi = \Omega_b$ (Athanassoula et al. 1983; Sparke & Sellwood 1987; Ceverino & Klypin 2007).

Orbits influenced by the bar through resonance interactions generally meet the condition that the orbital frequencies are commensurate with the bar's azimuthal frequency and its $m_b = 2$ mode (e.g. Ceverino & Klypin 2007). In the underlying theoretical analysis, of particular interest are planar orbits with bar resonances are the horizontal Lindblad resonances. These occur at specific intervals of the azimuthal frequency in the frame of the bar,

$$m(\Omega_\phi - \Omega_b) = \pm\kappa, \tag{1}$$

where $\kappa$ is the epicyclic or radial frequency. Analogous vertical resonances exist when replacing $\kappa$ with $\nu$ (Binney & Tremaine 1987).

For orbits where the azimuthal coefficient matches the bar mode, $m = m_b = 2$, we recover the inner and outer Lindblad resonances (ILR and OLR) of the bar (Lindblad 1963; Lin & Shu 1964). The horizontal ILR (hILR) is associated with the elongated *x1* orbits of the bar, located within the corotation radius. These orbits exhibit radial oscillations with two maxima in radius per orbital period, deviating from circularity, and are referred to as 'bar-supporting' orbits (Contopoulos & Papayannopoulos 1980; Athanassoula et al. 1983; Sparke & Sellwood 1987; Ceverino & Klypin 2007). The analogous vertical ILR (vILR) is associated with vertically extended orbits that oscillate above and below the disc plane, exhibiting two maxima per azimuthal period. This resonance emerges through a bifurcation of the *x1* orbit family, known as the *x1v1* family (Skokos, Patsis & Athanassoula 2002). For a more complete up-to-date census of the orbital 2:1 families, we refer the reader to Patsis & Athanassoula (2019). These orbits have characteristic *U*-shaped bends when viewed edge-on and are often referred to as 'banana orbits' (or BAN+ and BAN− as introduced by Pfenniger & Friedli 1991), which have since been thoroughly investigated within the context of the BPX (Combes et al. 1990; Athanassoula 1992; Patsis et al. 2002; Quillen 2002; Skokos et al. 2002; Martinez-Valpuesta et al. 2006; Quillen et al. 2014; Manos, Skokos & Patsis 2022).

The characterization of these resonances assumes axisymmetry and works best in the weak-bar limit; we depart from this idealized orbital framework and use the notation of radial frequency $\Omega_r$ and vertical frequency $\Omega_z$ in place of $\kappa$ and $\nu$ in this work. When the vertical ($\Omega_z$) and horizontal ($\Omega_r$) frequencies are nearly equal, the orbits are prone to vertical–horizontal resonance overlap (Combes et al. 1990; Pfenniger & Friedli 1991; Patsis et al. 2002; Skokos et al. 2002; Valluri et al. 2016). This occurs when either frequency becomes commensurate with the bar at an integer multiple of the azimuthal frequency of the orbit in the rotating bar frame. Since the term 'resonance overlap' generally takes into account resonance widths or spacings and requires analysis steps that emphasizes chaotic behaviour (as introduced in Chirikov 1979), which we do not undertake in this work, we describe the condition where the horizontal and vertical frequencies are nearly equal and the resulting susceptibility to resonance overlap as the intersection of the horizontal–vertical bar resonances.

Numerical studies of the populations of the BPX have characterized the stellar orbits to measure the role of bar resonances on the BPX in *N*-body models using orbit integration in a frozen potential (Portail, Wegg & Gerhard 2015; Abbott et al. 2017) and with a limited time window through time-evolving frequency analysis (Parul, Smirnov & Sotnikova 2020).

There has been substantial work focused on quantifying the differences in the BPX morphology (e.g. Patsis & Harsoula 2018; Smirnov & Sotnikova 2018) or BPX orbital populations within the context of varied galactic haloes, bulge concentrations, thin–thick disc combinations, or Hubble types (e.g. Quillen et al. 2014; Fragkoudi et al. 2017; Li et al. 2023, 2024; Ghosh et al. 2024; Tahmasebzadeh 2024). The vertically extended vILR orbits were initially proposed as the main component of the resonance-driven BPX structure but have since been found to account for only about ∼15 per cent of the BPX population, and to dominate just the outer radial extent of the feature with orbital families (e.g. 5:3 brezel in Portail et al. 2015 or 'warm ILR' in Beraldo e Silva et al. 2023) filling the central region (see also Valencia-Enríquez, Puerari & Chaves-Velasquez 2023; Zhang et al. 2024). This is consistent with what would be expected by the increase in radius of the resonance location in the context of a slowing bar.

Morphological characterization of the BPX with the inclusion of gas and on-going star formation (Baba, Kawata & Schönrich 2022), applying metallicities for direct comparison with the Milky Way (Debattista et al. 2020), and in cosmological contexts (Fragkoudi et al. 2020) has been more limited in scope.

Analysis of the role of resonance mechanisms of trapping or passage for producing the BPX feature started with the identification of radially overlapping resonant vertical and horizontal 2:1 orbits as introduced by Combes & Sanders (1981). Early work concluded that radial overlap of the resonances was maintained through the changes in surface density and scale height to maintain $\nu \sim \kappa$ (Combes et al. 1990). The resonance trapping BPX mechanism has since been studied in detail (Quillen 2002), finding bar-resonance passage elevating orbits through the v2:1 as the bar sweeps past orbits that are in the mid-plane of the disc, and without finding h2:1 and v2:1 necessarily occur at the same radius (Quillen et al. 2014). Both trapping and passage mechanisms have been shown in more recent work to successfully populate a BPX feature without any underlying gravitational instability (Sellwood & Gerhard 2020), and that resonance trapping and kinematic heating can occur simultaneously (Zozulia et al. 2024).

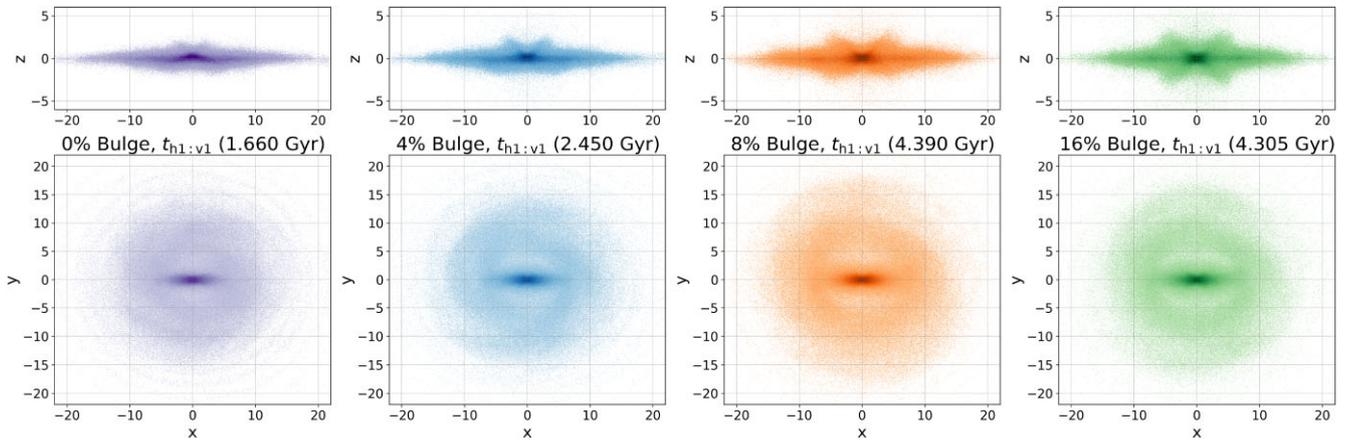

**Figure 1.** Each panel presents the overall density distribution from the simulations, shown in a frame fixed to the bar's inertial frame, with consistent density scaling across all frames. The models share the same disc mass but vary in bulge fractions, ranging from 0 per cent Bulge, 4 per cent Bulge, 8 per cent Bulge, 16 per cent Bulge, from left to right. The top panels show edge-on profiles, while the bottom panels display top-down views. All frames use the same fixed maximal colour saturation with a logarithmic scale to highlight surface density differences and non-axisymmetric features. Each galaxy in the suite has a strong bar, with outer spiral arm features, at the time. Variations in the central bulge mass result in bar morphology differences, especially in the central region. The selected time corresponds to the peak fraction of orbits with $\Omega_z = \Omega_r$ ($t_{h1:v1}$).

Most analysis of $N$-body simulations features at least one apparent bar-buckling event, which is the focus of the analysis in many other prior studies (e.g. Martinez-Valpuesta et al. 2006; Łokas 2019a, b); notably recent work by Li et al. (2023) outline a condition for buckling on the fraction of vertically extended bar-supporting orbits and find these stars cross the vertical 2:1 simultaneously with bar-buckling. Along with Quillen et al. (2014), they use this as evidence that bar-buckling plays a role in the horizontal-to-vertical resonance crossing and passage but emphasize that the resonance mechanism is then responsible for on-going BPX growth. This contrasts with recent findings by Sellwood & Gerhard (2020) who take a direct approach to characterizing the stellar orbital frequency components to focus on a comparative characterization of the BPX formation through buckling, resonance trapping with a slow bar, and resonance passage; they find stars do not need to be near the 2:1 resonance for their buckling instability to occur.

In this work, we study the effect of variations in the relative mass of a central classical bulge on the bar's final state and related bar-resonance processes. In this, we present evidence that the bar-resonance mechanism is sufficient to populate the BPX structure regardless bulge fraction but that the rate of bar-resonance BPX growth is moderated by the vertically extended orbit fraction, which is inherently impacted by the initial bulge mass fraction. We also characterize the evolution of the stellar bar, classical bulge, and BPX structure and their influence on each other throughout a galaxy's lifetime, as encoded in their observed properties and resonance evolution.

We outline our simulations and orbital frequency analysis techniques used to characterize the orbital components in Section 2. In Section 3, we provide a detailed analysis of how increasing the bulge fraction impacts the resulting bar and the bar-resonance populations that contribute to the BPX structure. In Section 4, we place our findings in the context of existing literature and discuss their broader implications. Finally, we summarize our conclusions in Section 5.

## 2 METHODS

### 2.1 Simulation parameters

The four $N$-body simulations presented here explore the impact of including a central stellar bulge on bar and BPX evolution in isolated galaxies. Each set of initial conditions was created with a modified version of MakeNewDisk (Springel, Di Matteo & Hernquist 2005) and evolved with the moving-mesh, finite-volume hydrodynamics and gravity code AREPO (Springel 2010) – though we only make use of gravity. The models have a typical three-component set-up: halo, stellar disc, and central bulge. First, we include a Hernquist (1990) dark matter halo with mass $10^{12}\,M_\odot$, $R_{200} = 163\,\mathrm{kpc}$, and concentration parameter $c = 11$. Secondly, each simulation includes a radially exponential and vertically isothermal stellar disc of mass $4.8 \times 10^{10}\,M_\odot$, radial scale length of $\sim 2.7\,\mathrm{kpc}$, and vertical scale height of $\sim 0.32\,\mathrm{kpc}$.[1] And thirdly, we include a stellar bulge with a Hernquist (1990) profile and scale length 3.15 kpc. We tested four bulge masses: 0, 2, 4, and $8 \times 10^9\,M_\odot$. These are 0 per cent, 4 per cent, 8 per cent, and 16 per cent of the disc mass shown in Fig 1 and throughout in purple, blue, orange, and green, respectively.

For the disc and bulge, we used a mass resolution of $6 \times 10^4\,M_\odot$, corresponding to about $8 \times 10^5$ disc particles. In the most massive bulge run, this resolution corresponds to about $1.3 \times 10^5$ bulge particles, with fewer particles in the lower mass runs commensurate with each lower bulge mass fraction. For the halo, we used a mass resolution of $3 \times 10^5\,M_\odot$, corresponding to about $3.2 \times 10^6$ particles. We used a softening length of 40 pc for the disc and bulge and 80 pc for the halo. To ensure our results are not resolution dependent, we ran simulations with the same increase in bulge fraction with $8\times$ lower mass resolution (and $2\times$ greater softenings) that resulted in the same qualitative behaviour for each model with regards to the final state of the galactic bar, the emergence of sudden asymmetry events, and the final morphology of the emergent BPX. Our choice of softening lengths is smaller than the typical Power et al. (2003) relation, but is closer to the softening used in resolved interstellar medium (ISM) models (e.g. Hopkins et al. 2018; Marinacci et al. 2019). A consistency check in which the softening lengths of a similar set-up were doubled was performed in Beane et al. (2023).

[1]The disc scale length changes slightly with different bulge masses. The disc scale height is always $0.12\times$ the scale length.

## 2.2 Stellar bar and BPX properties

Each simulated galaxy in our suite evolves a strong stellar bar and a prominent BPX well exceeding the disc scale height, as measured through the amplitude of the $m = 2$ mode and the vertical projection of that mode. To characterize the galactic bars of our simulation suite, we present the galactic bar amplitude as the strength of the $m = 2$ Fourier mode of the face-on surface stellar density, $A_2$, of each model galaxy normalized by the amplitude of average surface density, $A_0$,

$$A_{bar} = \frac{1}{N} \left| \sum_j e^{2i\phi_j} \right|, \quad (2)$$

over the $j$th stellar disc particles (Sellwood & Athanassoula 1986; Debattista et al. 2006; Baba et al. 2022). We compute $A_{bar}$ in cylindrical radial annuli of 0.25 kpc to determine the radius of the maximal $m = 2$ amplitude (bar-peak), the subsequent radius where the $m = 2$ feature drops to ~68 per cent of its maximum (bar-edge) (e.g. Athanassoula & Misiriotis 2002; Li et al. 2023), and to ~20 per cent of its maximum (bar-outer-limit) as used for the bar of the Milky Way (e.g. Wegg, Gerhard & Portail 2015). We show $A_{bar}$ within the bar-edge for each model over time as calculated within the disc in Fig. 2 with the middle panel time axis showing the bar-amplitude growth from the final time that $A_{bar} = 0$ plotted as a function of the number of bar rotations measured. For studying the bar evolution and studying its progression in comparison across the range of bulge fractions in our suite, we utilize the strong bar threshold of $A_{bar} = 0.2$ (Aguerri et al. 2001; Bland-Hawthorn et al. 2023) to align our initial bar states to enable comparison in further analysis, this is marked in the middle panel. The lower panel shows the $A_{bar}$ progression over time (in Gyr) after each model galaxy reaches the strong bar threshold.

The bar-pattern speed ($\Omega_{bar}$) is measured as the change in the $m = 2$ bar mode per recorded simulation time-step (snapshots are kept at 0.5 Myr intervals) as seen in the top panel of Fig. 2 with the shade ranging from light (slow) to dark (fast) colouring in all three panels. Before the bars reach the strong bar threshold, each has initial pattern speeds that oscillate between 45 and 60 km s$^{-1}$ kpc$^{-1}$ as the bar starts to form.

Each model in our suite results in a strong BPX feature aligned with the galactic bar with vertical extent exceeding the disc scale heights at amplitudes of ~1.5–1.7 kpc. The BPX amplitude is measured as the height of the $z$ stellar disc component above the bar region,

$$A_{BPX} = \sqrt{\frac{1}{N} \left| \sum_j z_j^2 \right|}. \quad (3)$$

This provides a measure of the apparent strength of the BPX region within the radial extent of the bar at each time in each simulation

(Athanassoula & Misiriotis 2002; Athanassoula 2008a), which we show in Fig. 3. This figure shows the similarity in each model's final BPX height, with only the 0 per cent Bulge model lagging slightly below the otherwise equal final amplitudes around 1.7 kpc. The progression in time is shown after the bar amplitude permanently breaches the strong bar threshold, and the plot colour saturation is set by measuring the asymmetry of the $z$-projection of the $m = 2$ Fourier mode of the stellar density. This value is also shown in the lower panel for reference. The progression of the BPX amplitude and height is steadier over time with increasing bulge fraction, with the most intense rates of increases in the amplitude occurring in the 0 per cent and 4 per cent Bulge models and with a steady rate of growth seen in the 16 per cent Bulge model.

Since the evolution of a BPX feature from a stellar bar can be associated with a sudden loss of vertical symmetry in the bared region of a stellar disc, we follow Athanassoula & Misiriotis (2002) and Baba et al. (2022) in measuring the asymmetry of the $m = 2$ mode's vertical extent as $A_{asym}$ over the full evolution of each galaxy in our suite,

$$A_{asym} = \frac{1}{N} \left| \sum_j z_j e^{2i\phi_j} \right|. \quad (4)$$

This is the value shown over time in the lower panel of Fig. 3 and used as colour-intensity scaling in later figures as it serves as a useful reference for the apparent symmetry of the disc at any time in each galaxy model. If the amplitude of that asymmetry is of the order of the simulation scale height, then it can be said that a significantly dramatic event generally referred to as ' bar-buckling' has occurred (Athanassoula 2003; Bournaud et al. 2005; Debattista et al. 2006; Athanassoula 2008b). We measure the velocity dispersion ratios throughout the suite at all times and all radii and find no instance where the ratio $\sigma_{vz}/\sigma_{vr} \lesssim 0.3$, a theoretical threshold established for the triggering of a firehose instability (e.g. Raha et al. 1991; Merritt & Sellwood 1994; Athanassoula 2008a). We acknowledge that this is not a comprehensive test of the mechanism underlying bar-buckling, but thorough investigation of buckling and buckling thresholds is not within the purview of this paper. To avoid confusion with sudden asymmetries driven by an underlying gravitational firehose instability event (as introduced by Raha et al. 1991), we explicitly avoid using this terminology in this work.

## 2.3 Orbital frequency analysis

To study the resonance structure evolution of our galactic discs, we perform frequency analysis on the evolving stellar orbits to measure the fraction of stars commensurate with the bar as it and its associated resonance features evolve. We opt for an orbit analysis technique to capture the frequencies and underlying commensurabilities as they change in the live potential as bar evolution is not often in the regime where idealized orbit analysis completely captures resonant behaviour. Our frequency analysis is also unique in that it is completed in dynamically set sliding time windows at all times, enabling us to characterize the nuances of an evolving potential without the implicit assumptions of true adiabatic evolution with a frozen potential and backward integration (e.g. Pfenniger & Friedli 1991; Quillen et al. 2014; Łokas 2019b). This is sensitive to the changing parameters of the stellar bar as resonances develop, drive the emergence of vertical structure, and trigger temporary asymmetry events. This technique results in more robust orbital characterizations compared to previous work that also performed frequency analysis in time-evolving orbits to identify resonances throughout galaxy models with diverse galactic parameters (e.g. Ceverino & Klypin 2007; Gajda, Łokas & Athanassoula 2016; Wang, Athanassoula & Mao 2016; Parul et al. 2020; Sellwood & Gerhard 2020; Baba et al. 2022; Li et al. 2023).

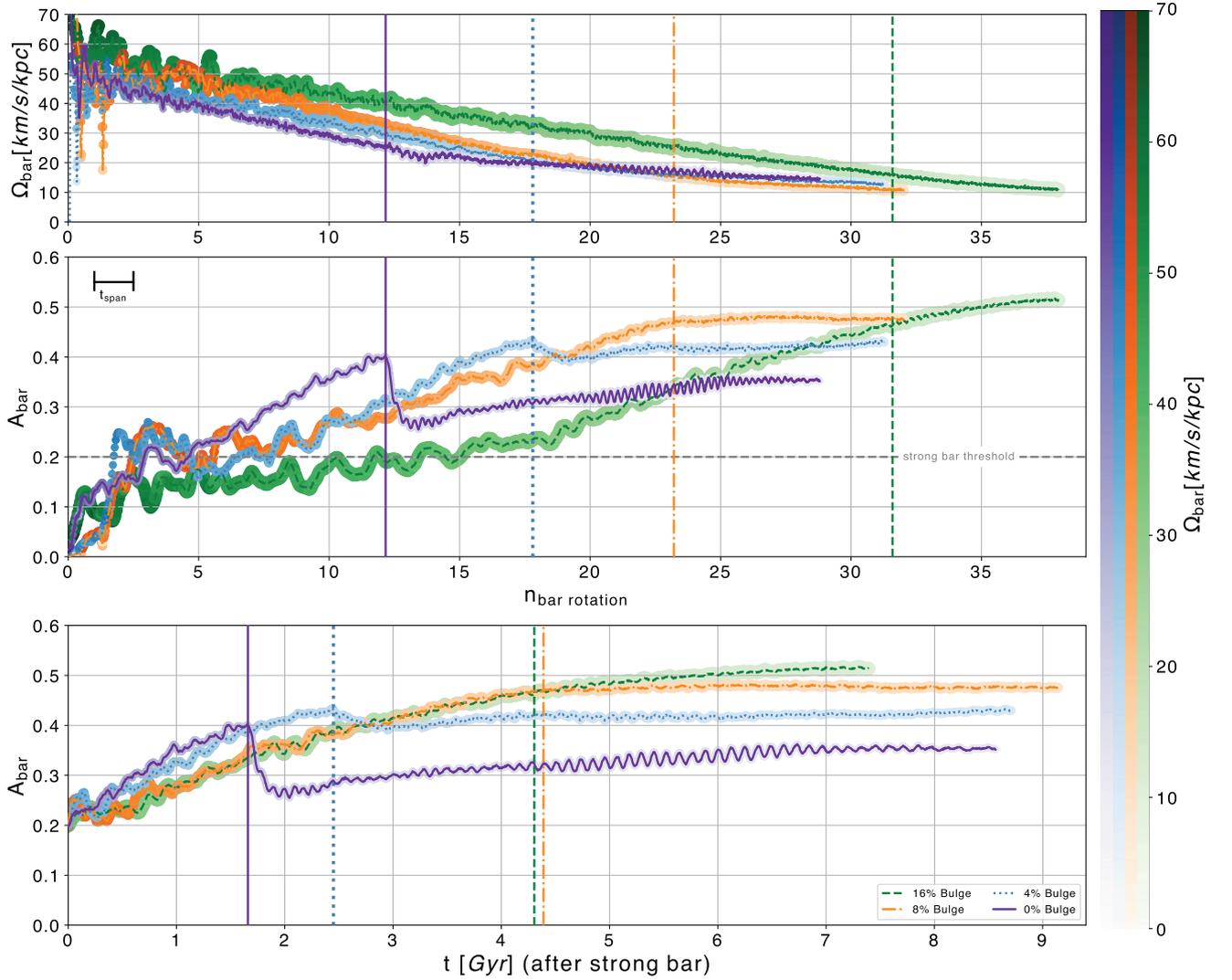

**Figure 2.** The bar-pattern speed and bar-amplitude of each model is displayed. The models include 0 per cent Bulge (solid line), 4 per cent Bulge (dotted line), 8 per cent Bulge (dashed–dot line), to 16 per cent Bulge (dashed line). Matching markers along the lines represent the bar-pattern speed ($\Omega_{\mathrm{bar}}$) with darker colours indicating faster speeds and lighter colours showing slower speeds as the bar approaches a steady state. Four vertical lines mark $r_{\mathrm{h1:v1}}$, the time-step when the fraction of stars with equal radial and vertical frequencies is maximal. *Top panel* shows the bar-pattern speed evolution from the time-step where $A_{\mathrm{bar}} = 0$, marking the point at which a measurable bar first forms in each model. The *x*-axis is expressed in bar rotations. *Middle panel* shows bar-amplitude evolution, also from where $A_{\mathrm{bar}} = 0$. For reference, the orbital frequency analysis window is marked $t_{\mathrm{span}}$ in the plot's upper left, spanning $1.5 \times 2\pi/\Omega_{\mathrm{bar}}$. *Bottom panel* shows the bar evolution over simulation time in Gyr with $t = 0$ set when the bar amplitude permanently exceeds the strong bar threshold of $A_{\mathrm{bar}} = 0.2$ (indicated by the horizontal grey dashed line in the middle panel).

We utilize `naif` frequency analysis code (Beraldo e Silva et al. 2023) to extract the underlying directional frequencies of the stellar trajectories at all times in the simulation for individual stellar orbits. We perform the analysis in rolling dynamic time windows of $t_{\mathrm{span}} = 1.5 \times 2\pi/\Omega_{\mathrm{bar}}$ where $2\pi/\Omega_{\mathrm{bar}}$ is the bar period (a full rotation over $2\pi$ of one end of the bar) at the centre of the time window. In the middle panel of Fig. 2, we include a marking of $t_{\mathrm{span}}$ for reference to the evolution of the bar amplitude and period over each simulation as the time axis of this plot is in number of bar rotations. With this technique, we recover the fundamental frequencies for each component of each stellar trajectory: $\Omega_r$, $\Omega_z$, in cylindrical coordinates, and $\Omega_x$ for which we subtract the bar angle from every individual star's azimuthal coordinate to align with $y = 0$ and calculate the $x$-frequency in this bar inertial frame. We opt for $\Omega_x$ in place of the azimuthal period with the bar period subtracted ($\Omega_\phi - \Omega_{\mathrm{bar}}$) as frequencies in the Cartesian frame of the bar have been shown to more cleanly separate orbital families (e.g. Valluri et al. 2016, appendix B). We further tested this with our frequency techniques for resonances in both the bar-subtracted azimuthal frame directly ($\Omega_\phi - \Omega_b$) and with $\Omega_x$ calculated within the inertial frame of the bar and (in agreement with prior work) found that the latter provided better separation of the azimuthal frequency populations, especially evident in distinguishing between kinds of corotators.

We are most interested in the families of orbits that undergo the bar-resonance interactions by selecting populations of stars with the relevant frequency ratios: radial 2:1 within $1.98 < \Omega_r/\Omega_x < 2.02$ (hereafter, h2:1) and vertical 2:1 as $1.98 < \Omega_z/\Omega_x < 2.02$ (hereafter, v2:1). In studies of the BPX feature, other families of orbits related to the 2:1 resonances have been found to contribute significant fractions of the population of stars in the BPX (Patsis et al. 2002; Skokos et al. 2002; Valluri et al. 2016). Of particular interest are vertically extended brezel orbits (Portail et al. 2015) or 'warm ILR cloud' orbits (Beraldo e Silva et al. 2023), so-called since they are analogous to crossing the bar vILR. This orbital family that exhibits frequency ratios in the range associated with passage through the 2:1 vertical-to-azimuthal in the bar frame as $\Omega_z/\Omega_x < 2$ (hereafter, xv2:1, to indicate crossing the v2:1 threshold; Abbott et al. 2017; Łokas 2019b; Parul et al. 2020; Beraldo e Silva et al. 2023). Characteristic orbits from each simulation as selected on these frequency ranges are shown in Fig 4 at the time corresponding to 1 Gyr before $t_{h1:v1}$ in each model with corresponding colours.

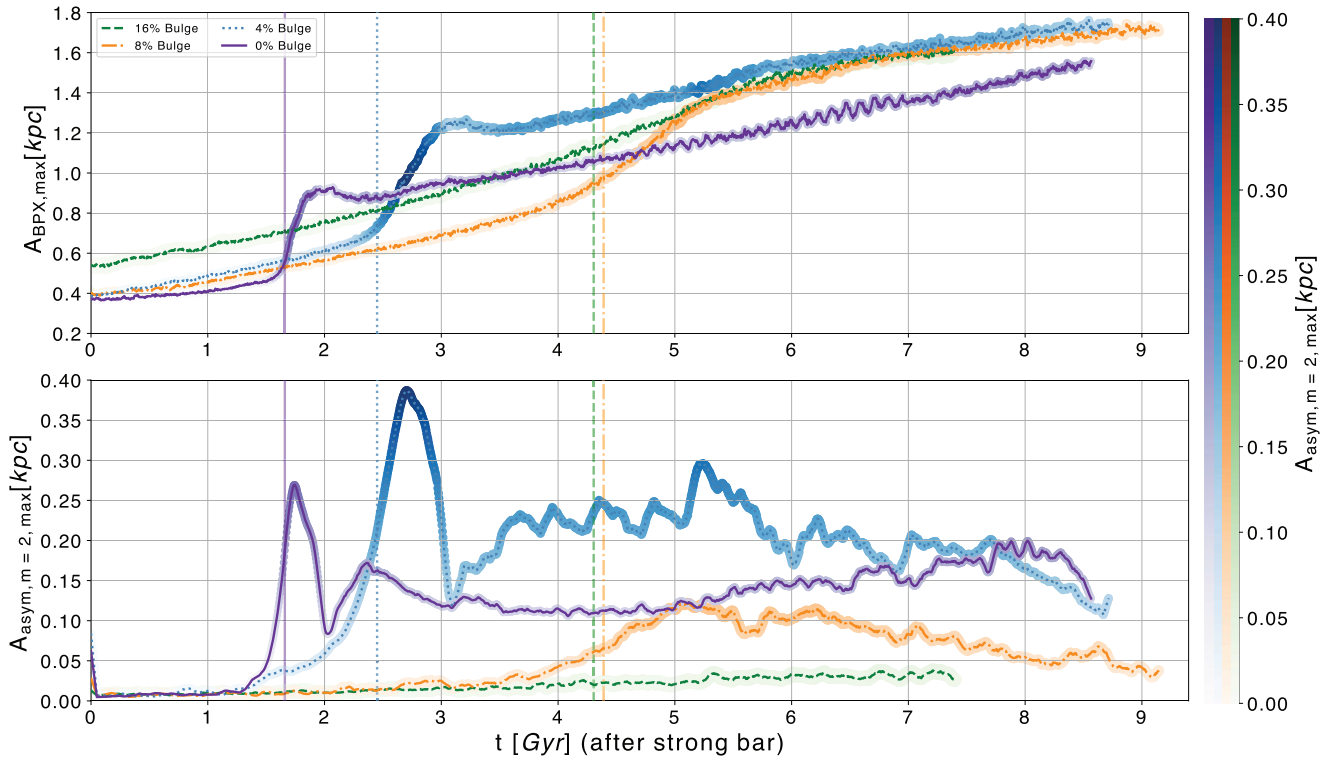

**Figure 3.** *Top panel* shows the BPX amplitude ($A_{BPX}$), defined as the maximal vertical projection of the $m = 2$ mode above the disc plane. The line styles for each model are consistent with previous figures. *Bottom panel* shows the vertical projection of the $m = 2$ asymmetry, $A_{asym}$ (equation 4), measured within the bar radius. High values (and dark saturation) correspond to brief periods of significant vertical asymmetry. Intensity or saturation accompanying lines are scaled to the amplitudes at each time as a reference. Each model includes a vertical line corresponding to $t_{h1:v1}$, the peak of the number of stars with equal radial and vertical frequencies.

We highlight one family of corotators in the left panel of these frequency orbit examples that show a pronounced spike in the $\Omega_r = \Omega_x$ with the distinct morphology of the trojans found at the bar L4 Lagrangian point (D'Onghia & Aguerri 2020), which are difficult to distinguish from other corotators in the purely cylindrical frame of analysis. The bar-supporting h2:1 orbital selection is very well constrained to the apparent bar region, as seen in the top-down view provided. The vertically extended v2:1 population exhibits the expected 'banana orbits' or $U$-bend shape and is aligned to the bar as seen edge-on and face-on in the selection. The final selection is on the frequency selection range identified to support the BPX structure requiring both brezel and bar-supporting frequency ranges within xv2:1 & h2:1, producing a top-down bar-view and BPX profile from edge-on view before the onset of BPX growth.

To show the evolution of the distribution of relevant orbital frequency ratios, we show for each simulation in Fig. 5 the time of peak population fraction of intersecting horizontal–vertical resonances ($t_{h1:v1}$) with distribution of $\Omega_r/\Omega_x$ and $\Omega_z/\Omega_x$ of all stellar particles. The colour in each point in the distribution is set by the average radius of particles with the corresponding combination of values of $\Omega_r/\Omega_x$ and $\Omega_z/\Omega_x$. The overall radial distribution is shown in the lower left insert of each plot along with the relevant colourbar scaling average radius shown in the main subplot. Contours that mark the density of the distribution of frequency ratios are overlaid in each plot with the corresponding line styles for each model to all other figures in the work.

On each subplot in Fig. 5, we mark the specific frequency range selections shown in Fig. 4 that correspond to the h2:1 (pink, star), v2:1 (salmon, diamond), xv2:1 (blue-grey, triangle), and one population of trojans corotators that correspond to a frequency feature of interest at $\Omega_r = \Omega_x$ (red, square). A distribution of bulge particles deviating from the overall distribution are visible in the $\Omega_z < 2\Omega_x$ range across a wide swath of $\Omega_r/\Omega_x$ with low radius and increasing intensity in the plots corresponding to the higher bulge fraction models. Each plot shows sharp spikes of the h2:1 population at $\Omega_r = 2\Omega_x$ and the v2:1 population at $\Omega_z = 2\Omega_x$, with a broader distribution of xv2:1 at $\Omega_z \leq 2\Omega_x$. There is only a low bump rather than a visible spike in $\Omega_z = 2\Omega_x$ for the 16 per cent Bulge model. The evolution of these distributions over time in each simulation are shown without the background radial in Fig. 6. They start relatively smoothly for both frequency ratios at early times in the evolution before each model reaches the strong bar threshold. By the time each reaches the strong bar threshold, there are sharp features (especially h2:1) in the $\Omega_r/\Omega_x$ distribution corresponding to the same features of Fig. 5, but the accompanying features in $\Omega_z/\Omega_x$ (v2:1 and xv2:1) do not emerge until the bar has had time to continue evolving and growing. We include the stacked corresponding figure for $t_{h1:v1}$ and, finally, a distribution at late times in the model around 7 Gyr after reaching the strong bar threshold. At these late times each simulation exhibits a dominant h2:1 feature and a broad xv2:1 feature, and none have a v2:1 spike. Each subplot has an inlaid radial distribution in the lower left panel and has corresponding step-plot histograms and overlaid density contours.

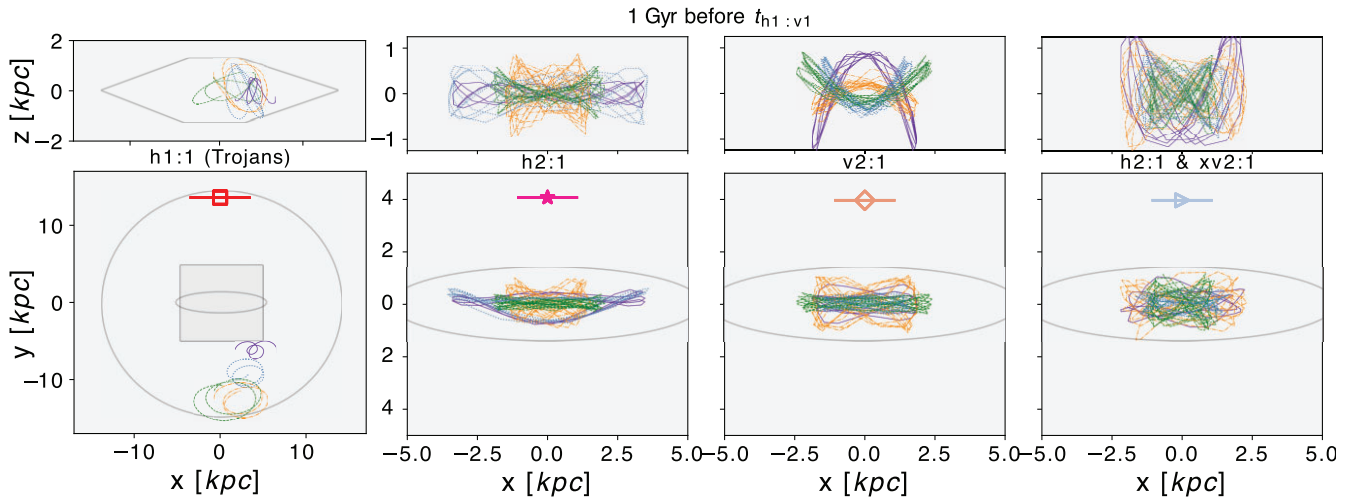

**Figure 4.** We display selected orbits in the bar's inertial frame aligned with $y = 0$ for the frequency ratio ranges of interest at a time 1 Gyr before $t_{h1:v1}$. *Top panel* shows the edge-on view and *bottom panel* shows a top-down view. Frequency ranges are marked on Figs 5 and 6 (from left to right): highlighted corotators (square, $\Omega_r/\Omega_x = 1$), radial-to-azimuthal bar-supporting h2:1 (star, $\Omega_r/\Omega_x = 2$), vertical-to-azimuthal v2:1 (diamond, $\Omega_z/\Omega_x = 2$), and bar-supporting brezel orbits with both h2:1 and xv2:1 (triangle, $\Omega_z/\Omega_x \leq 2$ and $\Omega_r/\Omega_x = 2$). The morphology of the orbits display libration around the Lagrangian points L4 and L5, confinement to the central bar region, the vertical $U$-bend banana structure, and the vertically extended brezel boxy-bar shape, respectively, for each category. One orbit from each model and category is shown, with line-style as in previous figures corresponding to each simulation model. Left-most frame shows light grey contours to mark the outer disc and central bar with an inset box corresponding to the dimensions of the other three frames that showcase the morphology of characteristic orbits of each frequency selection of interest.

## 3 RESULTS

In this work, we find the inclusion of a central galactic bulge alters the evolution of the stellar bar but that all models eventually have a central BPX. With a more massive the bulge fraction, the rate of bar-supporting orbits crossing the associated vertical resonance and BPX growth is steady. We also present the conditions necessary to prevent the resonance crossing rate from increasing dramatically, which circumvents a sudden loss of symmetry in the central bar region that coincides with an increase in the resonance crossing rate.

### 3.1 Bulge size impact on bar strength

We find that a bar in the presence of a substantial bulge evolves more steadily and ends up longer and stronger than a bar that evolves with a less massive bulge. This finding is in alignment with previous *N*-body studies (e.g. Athanassoula & Martinet 1980) and is supported by observations that find early-type disc galaxies with more massive bulges are accompanied by longer, stronger stellar bars (Elmegreen & Elmegreen 1985; Erwin 2005; Aguerri et al. 2009; Hoyle et al. 2011).

In all four models, the bar amplitudes exceed the threshold commonly used to define a strong bar, with $A_{bar} > 0.2$; the 0 per cent to 8 per cent Bulge models are strongly barred within the first $\sim$5 rotations, but the 16 per cent Bulge model does not breach the threshold permanently until after 15 bar rotations, for most of which it exhibits a weak-bar state. They all exhibit steady growth for at least 1.5 Gyr following this time. Notably, the 0 per cent Bulge and 4 per cent Bulge models each exhibit a decline in amplitude at $t_{h1:v1}$, which corresponds to increases in the rate of resonance passage and when the highest fraction of stars traverse horizontal–vertical resonances, as discussed in Section 3.3. The galaxy morphology is shown at this time in Fig. 1. The models with 8 per cent and 16 per cent bulge fractions show no noticeable decline in bar amplitude. Although the bar amplitude evolves similarly in strength during the first 1.5 Gyr of the strong bar phase in each model (to $A_{bar} \sim 0.4$), by the end of the simulations $A_{bar}$ stabilizes at higher final strengths as the bulge fraction increases.

Each bar slows during the first 2 Gyr after surpassing the strong bar threshold of $A_{bar} = 0.2$, decreasing over the next 6–7 Gyr as shown in Fig. 2. The 16 per cent Bulge model initially maintains it high early pattern speed; however, when it eventually does reach the strong bar threshold it is a slower bar than each of those in the lower bulge fraction models. The deceleration is well documented in *N*-body galaxy simulations and occurs as angular momentum is transferred to the live dark matter halo (e.g. Hernquist & Weinberg 1992; Debattista & Sellwood 2000; Athanassoula & Misiriotis 2002; D'Onghia & Aguerri 2020). We acknowledge that this slowing can be influenced or even halted by including the gas component (e.g. Athanassoula 2003; Athanassoula, Machado & Rodionov 2013; Beane et al. 2023), but addressing this is beyond the scope of this work.

### 3.2 Tendency for BPX emergence

Each galaxy model in our suite develops a prominent BPX feature with continuous growth as the bar-supporting stars cross through the vertical resonance. Including a classical bulge significantly impacts the stability of the rate at which orbits undergo this resonance passage.

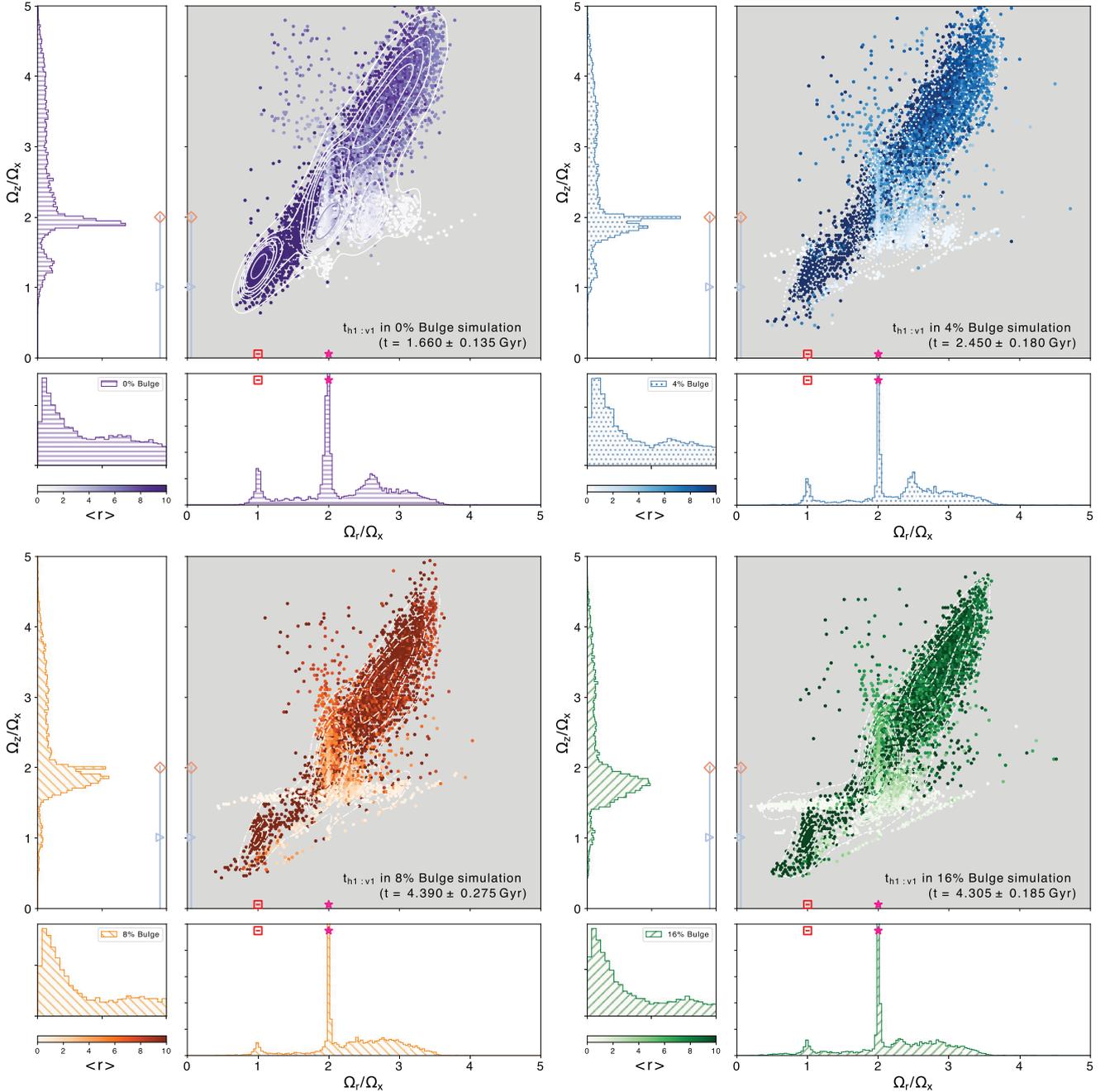

**Figure 5.** For each simulation, we present the frequency ratio distribution $\Omega_r/\Omega_x$ and $\Omega_z/\Omega_x$ centred on $t_{h1:v1}$, the time of peak $\Omega_z = \Omega_r$ fraction. Histograms of each distribution are shown along the $x$- and $y$-axes of the central plot, highlighting the fraction of orbits across different frequencies with peaks corresponding to bar-supporting populations. Contours correspond to the density, with line styles consistent across models. Bins are coloured by the mean radius of stars, increasing in saturation with increasing radius. The lower left inset shows the radial distribution at the central time. Frequency ratios indicating bar-resonance interactions are marked and correspond to the selected orbit examples in Fig. 4. Each model exhibits a distinct peak for the h2:1 radial-to-azimuthal bar-supporting population (star) and a broader peak for v2:1 stars transitioning through the vertical-to-azimuthal resonance (diamond), which extends into the xv2:1 region $\Omega_z/\Omega_x \leq 2$ (triangle). We also highlight a peak corresponding to stars at corotation with the bar, liberating around the Lagrangian points (square). In the simulations that include a classical bulge, a low radius, horizontally extended feature appears within $1.5 < \Omega_z/\Omega_x < 2$ contributed by the bulge from early in the simulations.

In the 0 per cent Bulge model, sharp growth in the BPX feature occurs at 1.7 Gyr, triggering a significant $m = 2$ asymmetry event soon after (see lower panel of Fig. 3). A 25 per cent drop in bar amplitude accompanies this event (see Fig. 2) and an asymmetry event of $A_{\text{asym}} \sim 0.27$ kpc, which is of the order of the scale height of the simulation. Similarly, in the 4 per cent Bulge model, a distinct increase in the BPX feature occurs at 2.5 Gyr, coinciding with an $m = 2$ vertical asymmetry event. Here, the BPX amplitude $A_{\text{BPX}}$ reaches $\sim 0.38$ kpc; there is an accompanying minor (less than 10 per cent) decrease in bar amplitude. Both events occur directly after $t_{h1:v1}$ and despite the sudden loss of symmetry in the bar, neither simulation at any radius around that time has velocity dispersion ratios that indicate an underlying causal gravitational instability mechanism. In this instance, the observed asymmetry is driven by the rapid resonance crossing occurring.

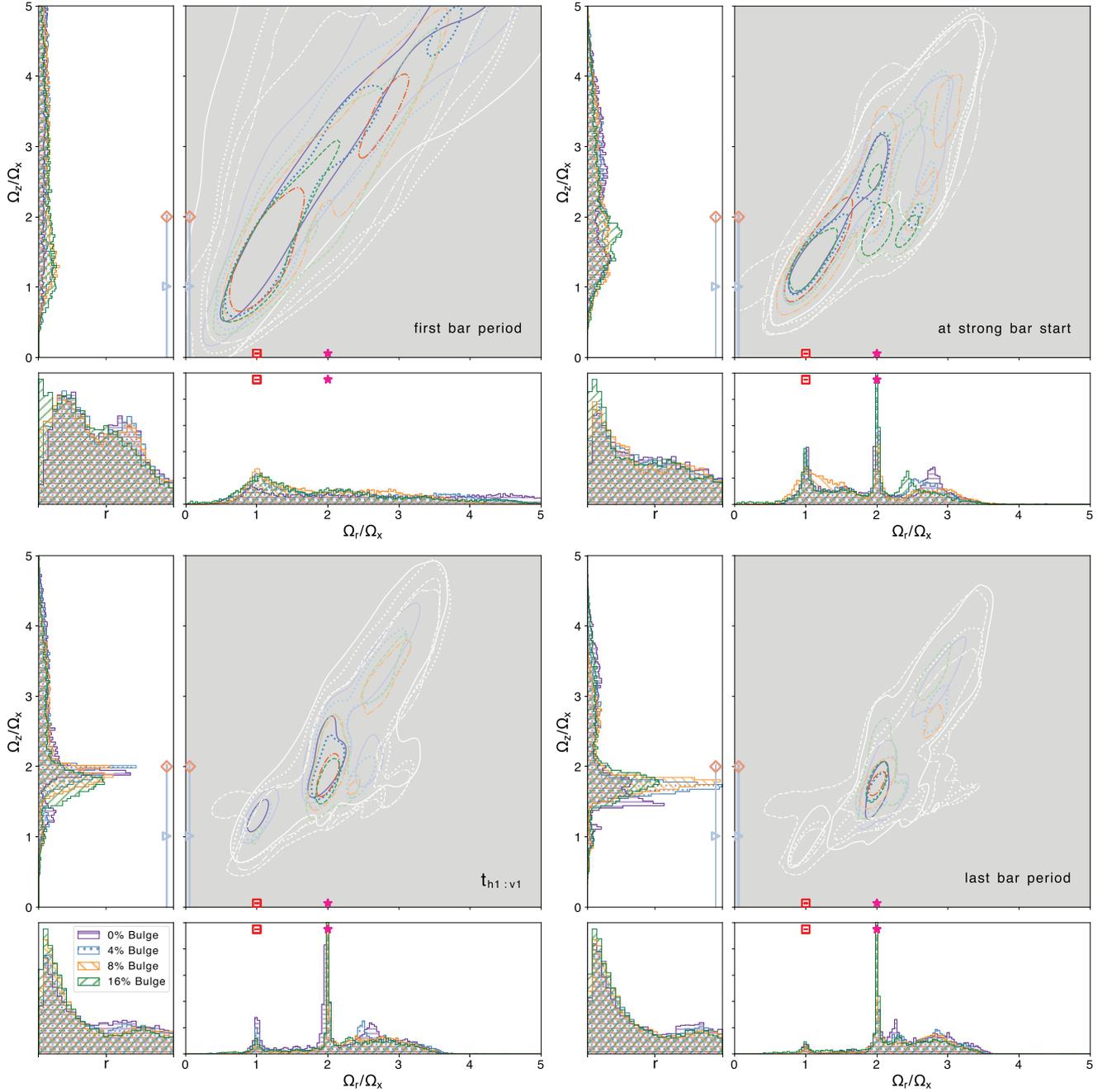

**Figure 6.** The panels (clockwise from top left) show the earliest time a bar is resolved, the time each model reaches the strong bar threshold, $t_{h1:v1}$ with the peak fraction of $\Omega_r = \Omega_z$ as in Fig. 5, and the last resolved bar period centred between 7 and 9 Gyr after reaching the strong bar threshold. This evolution shows the emergence and increasing fraction of stellar populations in bar resonances. Distributions of each simulation orbital frequency ratios $\Omega_r/\Omega_x$ and $\Omega_z/\Omega_x$ are shown with matching histogram hatch styles as in Fig. 5. Lower left insets show the radial distribution of each model at the central time. Each plot includes markings corresponding to key frequency ratio populations: h2:1, v2:1, xv2:1, and the featured corotators, shown in Fig. 4.

At later times in the evolution of the 8 per cent Bulge model, $A_{asym}$ gradually increases in the outer regions of the bar but the maximum amplitude remains lower than the disc scale height and persists for longer. Therefore, we do not classify this as the same kind of asymmetry event as in the 0 per cent Bulge and 4 per cent Bulge models. In general, as the bulge fraction increases, the galaxies do not exhibit the short-lived $m = 2$ vertical asymmetry events nor sudden drops in bar amplitude seen in lower bulge models.

This suggests that a substantial classical bulge prevents the bar from experiencing internally driven asymmetry and facilitates steady BPX growth. While the models with varying bulge fractions show differences in the evolution of $A_{BPX}$, they ultimately converge to similar amplitudes after the full evolution, with the BPX growth driven by bar-resonance interactions, as we discuss below.

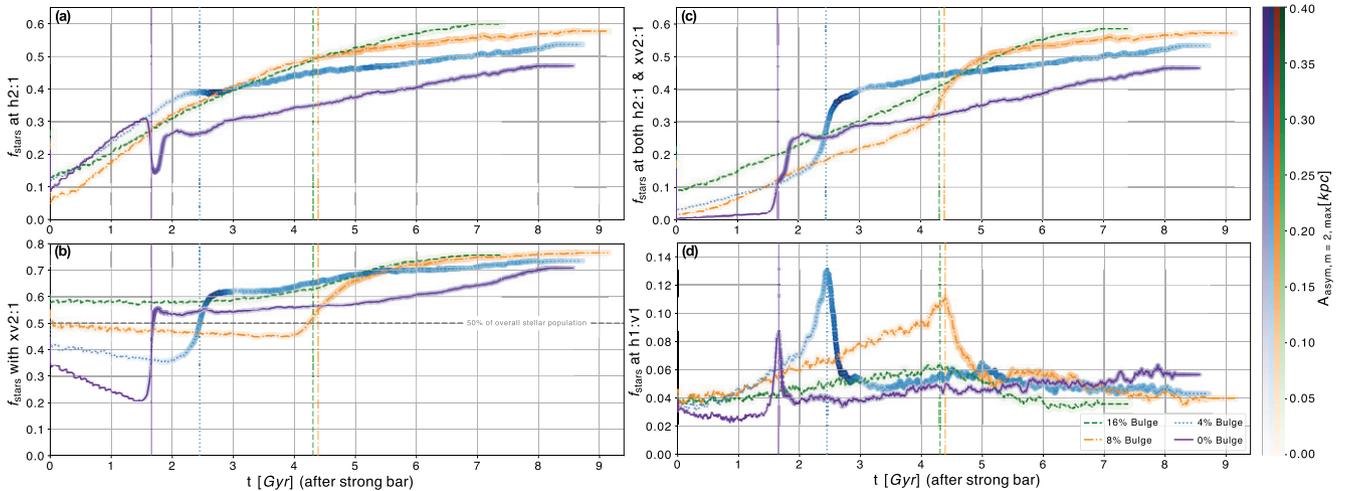

**Figure 7.** Each subplot shows the fraction of stellar orbits with specific frequency ratios throughout the evolution, starting when each simulation reaches the strong bar threshold. The saturation accompanying each line in each plot represents the $m = 2$ vertical asymmetry, consistent with the lower panel of Fig. 3, with the line styles used for each model throughout this work. (a) The overall fraction of h2:1 bar-supporting stars. The increase in bar-supporting orbits correlates with the rise in bar amplitude, as seen in Fig. 2. In the 0 per cent Bulge model, a temporary dip at $t_{h1:v1}$ corresponds to a substantial drop in the bar amplitude. (b) Fraction of stellar orbits exhibiting the vertically extended brezel xv2:1 orbits. Most orbits contributing to each increase in this fraction pass through the v2:1 resonance. Each rise in fraction of xv2:1 orbits coincides with a period of BPX amplitude growth, as seen in Fig. 3. (c) Steady growth of the stellar population with both h2:1 bar-supporting and vertically extended xv2:1 brezel frequency ratios. The majority of these stars previously had $\Omega_z / \Omega_x > 2$ and any increase is closely tied to the growth of the BPX feature. By the end of each model's evolution, 45–60 per cent of stars have both bar-supporting and brezel frequency ratios. (d) Fraction of stellar orbits in each model with approximately equal vertical ($\Omega_z$) and horizontal ($\Omega_r$) frequencies. These stars exhibit intersecting vertical and horizontal resonances if either frequency is commensurate with the bar. Vertical lines indicate the time $t_{h1:v1}$, marking the peak in each model, as shown in other figures.

## 3.3 BPX growth through resonance passage

The formation of a BPX occurs as h2:1 bar-supporting planar orbits at the horizontal bar resonance ($\Omega_r = 2\Omega_x$) evolve by crossing through the corresponding vertical v2:1 resonance ($\Omega_z = 2\Omega_x$) to become vertically extended xv2:1 orbits ($\Omega_z \leq 2\Omega_x$).

As introduced in Section 2.3, we identify each orbital family of interest by computing their frequency ratios. Each stellar orbit directional frequency is calculated over $1.5\times$ the bar period window ($2\pi/\Omega_{bar}$), for horizontal $\Omega_r$, vertical $\Omega_z$, and azimuthal $\Omega_x$ as calculated in the bar inertial frame. In Fig. 7, we show the fraction of stars at all times in the simulation that meet specific frequency ratios: (a) the bar-supporting h2:1 stellar fraction, (b) the brezel xv2:1 stellar fraction, (c) the stellar fraction with both bar-supporting h2:1 and xv2:1 brezel frequency ratios, and (d) $\Omega_r = \Omega_z$ (h1:v1). The peaks in the fraction of stars with $\Omega_r = \Omega_z$ in Fig. 7(d) correspond to $t_{h1:v1}$ for each model, marked by appropriate vertical line at that time. The growth rate of bar amplitude (Fig. 2) correlates with the increasing fraction of h2:1 orbits (Fig. 7a), while the growth of the BPX amplitude (Fig. 3) correlates with the fraction of stars with both h2:1 and xv2:1 (Fig. 7c). This suggests that the stellar bar drives the growth of BPX.

The sudden increase in stars with both h2:1 and xv2:1 orbits, shown in Fig. 7(c) for the 0–8 per cent Bulge models corresponds to the build-up of stars undergoing vertical–horizontal resonance intersection in Fig. 7(d) at $t_{h1:v1}$. The 16 per cent Bulge model, by contrast, shows a steady increase in the fraction of bar-supporting and brezel orbits, resulting in gradual BPX growth.

At $t_{h1:v1}$ both the 0 per cent and 4 per cent Bulge models exhibit a brief disruption in the bar evolution, indicated by a dip in $A_{bar}$ at 1.66 and 2.45 Gyr, respectively. This drop coincides with the peak in horizontal–vertical resonance intersection, possibly tracing bar-resonance overlap (Fig. 7d). The peak horizontal–vertical resonance intersection occurs just before sharp spikes in the vertical asymmetry of the $m = 2$ component ($A_{asym}$, see lower panel of Fig. 3), often identified as the signature of bar-buckling in galaxy simulations. This result differs from recent work by Li et al. (2023) that found the buckling was simultaneous with the bulk vertical 2:1 resonance crossing.

Overlap in the radial distribution of the orbits at horizontal and vertical bar resonances throughout the disc coincides with growth of the BPX. In Fig. 8, we plot the radial distributions of the h2:1 (thick lines) and v2:1 (thin lines) stellar populations, shading the radial range that corresponds to the middle three quartiles: $50 \pm 25$ per cent. The plot shows the h2:1 population consistently within the bar peak (the lower bound of the darker grey diagonal hatched region), while the v2:1 population shifts inward until both central radii align. The central radii remain overlapping until just before $t_{h1:v1}$ when the spike in the fraction of $\Omega_z = \Omega_r$ orbits ends. In contrast, the 16 per cent Bulge model never shows complete overlap between the central radii of the v2:1 and h2:1 populations and does not exhibit a sharp spike in the fraction of $\Omega_z = \Omega_r$ orbits, although their distributions overlap for most of the simulation. The time the radii distributions no longer overlap corresponds to when the fraction of orbits with $\Omega_z = \Omega_r$ returns to its initial value, implying that physical overlap of the h2:1 and v2:1 distributions successfully traces some aspect of bar-resonance overlap. For the majority of the simulation within the selected radial ranges of the h2:1 and v2:1 orbits, there is significant overlap in each model. When this overlap ceases, BPX growth slows and eventually stops, indicating that vertical–horizontal bar-resonance overlap is necessary for BPX growth.

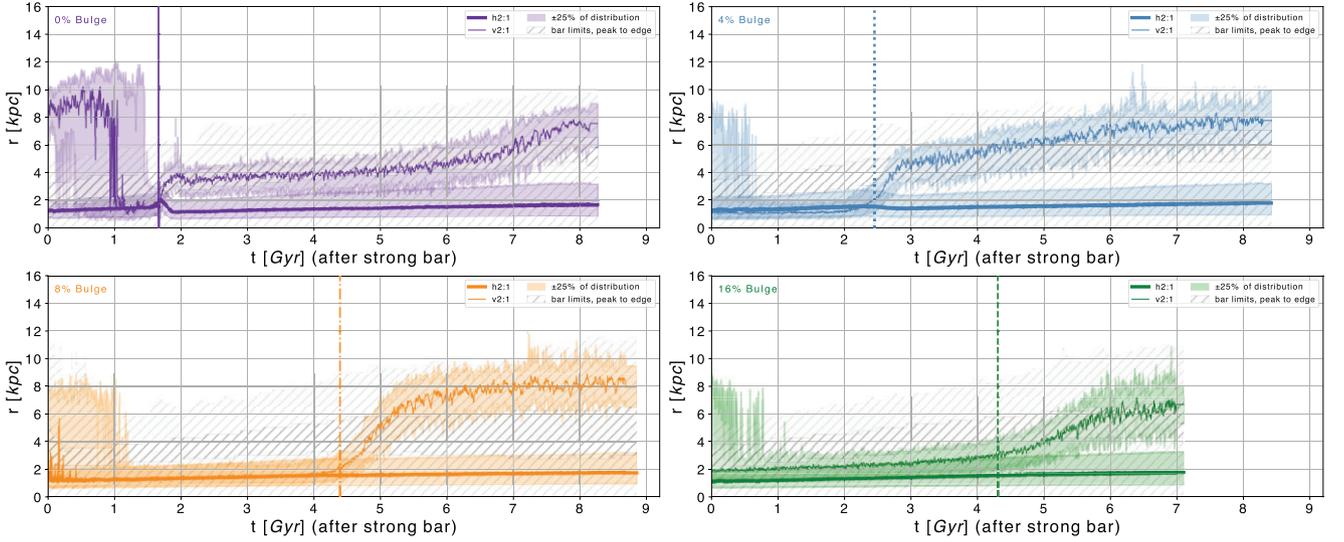

**Figure 8.** For each model, we plot the middle three quartiles ($50 \pm 25$ per cent) with shaded bounds of the radial distribution of orbits at the horizontal h2:1 (thick line) and vertical v2:1 (thin line) bar resonances over time. The bar's radial extent is shown by diagonal hatching with darker hatching from the radius of peak amplitude to the bar edge, where $A_{bar}$ drops to 68 per cent of its peak, and then continuing lighter hatching to the outer limit where $A_{bar}$ drops to 20 per cent. Vertical lines mark $t_{h1:v1}$, the point of peak intersection between vertical and horizontal resonances. In the 0 per cent Bulge model, h2:1 and v2:1 mid-quartile radii briefly converge at this peak. The 4 per cent and 8 per cent Bulge models show extended periods (2–3 Gyr) of convergence during the leading edge of the $t_{h1:v1}$ spike. At late times, the bar edge aligns with the v2:1 distribution, while the h2:1 distribution remains more centrally concentrated, always within the bar-peak radius.

## 3.4 Necessary conditions for steady BPX growth

In this work, we find that BPX feature formation occurs primarily when bar stars (specifically h2:1) cross through the associated vertical bar resonance (v2:1). This process occurs when the radial and vertical resonances intersect but we do not find that orbits undergoing this process are trapped in the vertical resonance for a prolonged period of time. We find that the efficiency of that resonance crossing process depends on the overall frequency distribution. The transition of bar-supporting orbits to increased vertical extension is a continuous process that intensifies as the bar grows; it becomes particularly efficient when the fraction of stars with $\Omega_z \leq 2\Omega_x$ drops below 50 per cent, leading to a sudden increase in the number of orbits primed for intersection of the vertical–horizontal resonances ($\Omega_z = \Omega_r$). This continues until the fraction of orbits with $\Omega_z \leq 2\Omega_x$ (specifically as contributed by the stellar disc) stabilizes above 50 per cent.

A significant fraction of stellar orbits in resonance with the bar do not need to remain in sustained resonance intersection (trapped at both the vertical and horizontal 2:1 resonance) for BPX growth to occur. Instead, stellar orbits with the horizontal 2:1 frequency ratio only exhibit the corresponding vertical 2:1 ratio for a brief period, indicating that resonance passage, rather than sustained resonance trapping drives BPX growth. Our findings show that, on average across all our models, bar stars with h2:1 frequencies are only in vertical 2:1 resonance ∼$4 \pm 10$ per cent of the time that they are h2:1. In contrast, stars with h2:1 frequencies spend more than $75 \pm 20$ per cent of their time in h2:1 resonance also as xv2:1 brezel orbits. This suggests that bar-supporting stars quickly cross the vertical 2:1 resonance and maintain their planar h2:1 frequency once they reach the xv2:1 brezel range. As the bar grows, more stars with the horizontal 2:1 frequency pass through the vertical 2:1 resonance, increasing the number of stars with bar-supporting h2:1 and brezel xv2:1 frequencies, as shown in the population fractions in Fig. 7(c).

It is evident that a sharp spike in $A_{asym}$ occurs following a sharp increase in the fraction of stars with $\Omega_r = \Omega_z$, seen in the 0 per cent and 4 per cent Bulge models of Fig. 7(d). In the 8 per cent Bulge model, there is a gradual increase in the $\Omega_r = \Omega_z$ fraction leading to a smaller spike in $A_{asym}$ in regions further out along the bar. In Fig. 9, we show the fraction of stars contributed by the disc with xv2:1 frequency ratios over time. The asymmetry event and increase in the $\Omega_r = \Omega_z$ fraction in each model continues until the fraction of xv2:1 brezel stars from the disc reaches 50 per cent, seen in the intersection of each $t_{h1:v1}$ vertical marking with the 50 per cent threshold in Fig. 9. In all models, the initial fraction of the overall xv2:1 stellar population contributed by disc stars is between 35 per cent and 45 per cent. There is a steady decline in the overall number of stars and the number of disc stars below this threshold except in the 16 per cent Bulge model. The decline ends with a spike in the fraction of stars at $\Omega_r = \Omega_z$, at time $t_{h1:v1}$. Once this threshold is exceeded, the resonance intersection spike of relatively high $\Omega_r = \Omega_z$ fraction (see Fig. 7d) ends. The increase in stars at $\Omega_r = \Omega_z$ can continue even after the overall stellar fraction stabilizes above 50 per cent. In the 16 per cent Bulge model, the overall fraction of stars with $\Omega_z / \Omega_x \leq 2$ never falls below 50 per cent, as seen in Fig. 7(b). The significant bulge prevents any sudden increase in fraction of orbits with equal vertical and radial frequencies, meaning there is not a bulk resonance intersection event and also no corresponding vertical asymmetry event.

Notably, the 0 per cent Bulge simulation exhibits a sudden temporary drop in the h2:1 bar-supporting stellar fraction (see Fig. 7a) after the fraction of stars with equal horizontal and vertical frequencies begins to rise sharply. The inclusion of a bulge in the other models seems to stabilize against this behaviour

## 4 DISCUSSION

Our results show that the inclusion of a classical bulge alters the distribution of stellar orbits from the start and stabilizes the growth of the fraction of stars supporting the bar. This work offers a novel examination of stellar frequency distributions as stellar bars and associated structures evolve. We use frequency analysis techniques to track orbits over time, identifying those undergoing resonance interactions with the bar. By characterizing frequencies throughout the simulation, we capture the evolution of a disc galaxy's live potential, focusing on the distribution of stellar resonances and the resulting resonance structures.

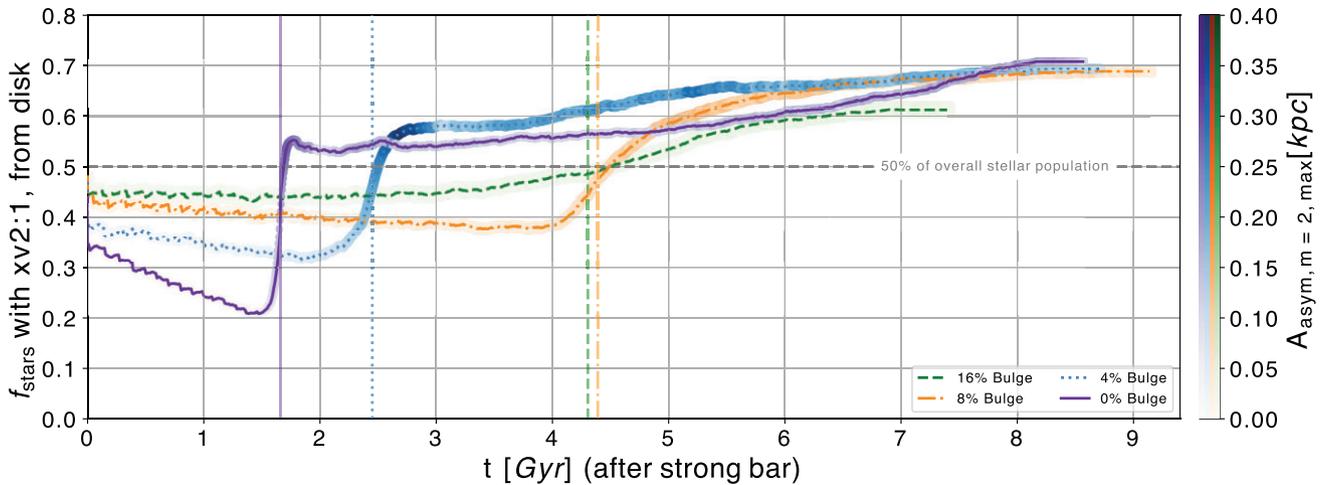

**Figure 9.** Overall stellar fraction of xv2:1 brezel orbits with frequencies below the vertical 2:1 bar-resonance threshold ($\Omega_z < 2\Omega_x$), specifically as contributed by the disc population. The initial fraction from disc stars with $\Omega_z/\Omega_x \leq 2$ ranges from 38 per cent to 45 per cent across all models. Simulations with bulge fraction of 8 per cent or lower show a steady decline in xv2:1 orbits during early evolution, with a more rapid decline as the bulge fraction decreases. Vertical lines mark $t_{h1:v1}$ at peak fraction of stars with equal horizontal and vertical frequencies, corresponding to peak horizontal–vertical resonance intersection. These lines intersect where the fraction of disc stars reaches 50 per cent, as this stabilizes the bar and BPX resonance evolution.

We find that if the fraction of stars with $\Omega_z \leq 2\Omega_x$ falls below 50 per cent, there will be an increase in stellar orbits with equal vertical and radial frequencies, indicating intersecting resonances if these orbits are also commensurate with the bar. During this time, the stellar bar population passes most of its stars through the available vertical bar resonance until reaching the 50 per cent threshold. Building on pathways of resonance trapping and passage that populate the BPX structure (e.g. Quillen 2002; Quillen et al. 2014), recent work by Li et al. (2023) proposed that a coherent bar-instability response will occur when there is a population of orbits with $\Omega_z > 2\Omega_x$ due to self-gravity of the stars in the bar, while a population with $\Omega_z/\Omega_x \leq 2$ stabilizes the bar against this response. Our results reveal that maintaining at least 50 per cent of the stellar population with $\Omega_z/\Omega_x \leq 2$ is necessary to prevent the sudden increase in bar asymmetry (measured here as $A_{asym}$). When this threshold is not met, frequency overlap between the vertical and horizontal components builds, leading to temporary vertical asymmetry in the bar region as more h2:1 bar-supporting stars pass through the v2:1 resonance. The presence of a significant bulge can keep the fraction of stars with $\Omega_z/\Omega_x \leq 2$ above 50 per cent, stabilizing the rate of resonance passage and preventing a spike in the fraction of orbits with $\Omega_z = \Omega_r$. This results in steady bar and BPX growth, as exhibited by our 16 per cent Bulge model and to a lesser extent in our 8 per cent Bulge model.

Our results further prove that BPX growth occurs through resonance passage as disc stars pass from horizontal 2:1 through vertical 2:1 bar resonances, increasing the fraction of bar-supporting stellar populations in the BPX structure. This adds to the literature showing resonance passage as a sufficient mechanism for BPX growth (Quillen et al. 2014; Sellwood & Gerhard 2020; Zozulia et al. 2024), though measuring resonant strength and associated timescales are beyond the scope of this work.

We find the measured resonance intersection is a transient phenomenon, with stars spending only a short time exhibiting both v2:1 and h2:1 frequencies. Once stars cross the vertical 2:1 threshold, their transition to xv2:1 is permanent, and they do not revert. Throughout this process, the stars maintain their bar-supporting horizontal 2:1 frequencies, indicating that BPX formation is a vertical extension of the bar rather than a disruption of it.

Low-order Hamiltonian analysis has been used to characterize the resonance evolution (e.g. Quillen et al. 2014) to illuminate the theoretical mechanisms that underlie the observed orbital evolution and to tie the resonance locations to a tight constraint on mid-plane mass density in a galactic disc, and to constrain bar history. This earlier work uses a bar-buckling event as a forcing mechanism to excite the resonance, but not all of our models have an asymmetry event. The development of higher order, more comprehensive theoretical models that allow for variation in the radial degree of freedom in the Hamiltonian analysis could account for both the horizontal and vertical terms simultaneously, but is beyond the scope of this work.

Our scenario for the resonance passage during bar growth supports an inside-out formation route for BPX growth, where the metallicity of inner BPX stars reflects the earlier stage of bar evolution. This would compound with a possible vertical metallicity gradient inherited from the distribution of stars in the bar before the BPX forms as found in Debattista et al. (2017) and (Fragkoudi et al. 2018). A sudden increase in resonance intersection corresponds with a sharp rise in BPX amplitude and would link observed metallicity spikes in BPX stars to prior bar destabilization events (e.g. Fragkoudi et al. 2020; Baba et al. 2022).

If bar growth is adiabatic, then the distribution of stars is heated by the resonances as it moves outward forcing orbits from the mid-plane into high inclination, leaving the resonance while at high inclination but the addition of a substantial quantity of gas through accretion could stall out the radial expansion of the BPX by shifting the vertical 2:1 resonance independent of the bar-growth rate (Quillen et al. 2014). More detailed study of the correlations between the rate of the slowing of the bar and the steadiness of the BPX growth could constrain observational tracers of the BPX and bar-galactic history or even reveal a history of substantial gas accretion events.

We find that the inclusion of a central bulge stabilizes the stellar bar as it forms and evolves, resulting in a longer and stronger bar with an increasing bulge fraction. Our findings suggest that galaxies with sudden BPX growth likely had minimal classical bulges, as larger bulges would instead facilitate steady BPX growth throughout bar evolution. Angular momentum exchange between the stellar bar, the dark matter halo, the gas, and bulge impacts the bar's steady state. Our results show that larger bulges contribute significantly to a longer and stronger bar, underscoring the importance of angular momentum exchange with the bulge (e.g. Athanassoula 2003). Recent cosmological simulation results indicate that the strongest predictor of BPX formation in disc galaxies is in the length and strength of the bar (Anderson et al. 2024). Since bars grow in length and strength over time, this finding from cosmological simulations with our finding that each of our models evolves a BPX eventually leads us to speculate that only relatively recently formed bars should not have a vertical extension in the form of a BPX.

Since we find that galaxies formed with early substantial classical bulges tend to have longer and stronger bars, the presence of a bulge is implicitly tied to the expected incidence of BPX in the local Universe. The impact of gas and dark matter halo components on angular momentum exchange will be addressed in future work.

In each simulation, an accompanying BPX structure forms as the fraction of bar-supporting orbits that are vertically extended beyond the disc steadily grows. Within 5 Gyr of the bar reaching the strong bar threshold, a substantial BPX structure emerges, regardless of including a bulge or bar stability over the formation period.

## 5 CONCLUSIONS

This study investigates the impact of an initial galactic bulge on the evolution of bars and BPX structures in disc galaxies over long isolated periods, with a live dark matter halo. We quantify how increasing bulge fractions affect the emergence and evolution of the bar and BPX using frequency analysis in the live potential. Our key findings are as follows:

(i) More massive bulges facilitate the formation of stronger, longer bars.

(ii) The inclusion of a bulge stabilizes the bar against asymmetry events such as short-lived $m = 2$ vertical asymmetry and sudden drops in the bar amplitude. The bulge plays a critical role in stabilizing the resonance evolution of the disc's stellar population, maintaining a majority fraction of stars with $\Omega_z \leq 2\Omega_x$. This condition prevents the surge in stars with $\Omega_z = \Omega_r$, which precedes a vertical asymmetry and significant drop in the bar amplitude.

(iii) The dominant orbital family that provides long-term support for the BPX feature is bar-supporting brezel orbits (stars with both $\Omega_r = 2\Omega_x$ and $\Omega_z \leq 2\Omega_x$).

(iv) BPX feature growth is driven through resonance passage of horizontal 2:1 stars crossing the vertical 2:1 resonance with a short time of $\Omega_z = \Omega_r$. Formation of the BPX does not require a sudden surge or prolonged trapping of horizontal 2:1 stars at vertical 2:1 bar resonance.

(v) This process proceeds steadily in the presence of a stabilizing bulge and more efficiently when there is radial overlap of bar-supporting stars exhibiting vertical bar resonance. The end of BPX growth occurs when the radial ranges of the vertical and horizontal 2:1 resonances cease to overlap.

## ACKNOWLEDGEMENTS

The authors thank the Flatiron Institute's Center for Computational Astrophysics and the Big Apple Dynamics school as the foundational hosts of this work. We would like to thank the referee, M. D. Weinberg, L. Beraldo e Silva, S. Lucchini, P. Patsis, and E. Łokas for helpful discussion in pursuit of this work and of the tools used throughout. We also acknowledge that this work was completed on the occupied ancestral home of the Ho-Chunk and 11 other nations of the land called Teejop (day-JOPE) since time immemorial. Contributions by RLM to this material were supported by the Wisconsin Space Grant Consortium under NASA Award No. 80NSSC20M0123, the Ruth Dickie Endowment of the UW-Madison Beta Chapter of SDE/GWIS, and the National Science Foundation Graduate Research Fellowship under Grant No. 2137424. Any opinions, findings, and conclusions or recommendations expressed in this material are those of the author(s) and do not necessarily reflect the views of the National Science Foundation.

## DATA AVAILABILITY

Simulations used in this analysis will be made available upon reasonable request. The following codes were used in the analysis and production of this paper: MATPLOTLIB (Hunter 2007), NUMPY (Harris et al. 2020), and SCIPY (Virtanen et al. 2020).

This paper has been typeset from a TeX/LaTeX file prepared by the author.